\documentclass[aps,prb,twocolumn,groupedaddress]{revtex4-1}
\usepackage{color}
\usepackage{dcolumn}
\usepackage{graphicx}
\usepackage{amsmath}
\usepackage{amssymb}

\begin{document}


\title{Critical behavior of dissipative two-dimensional spin lattices}


\author{R. Rota$^1$, F. Storme$^1$, N. Bartolo$^1$, R. Fazio$^{2,3}$, C. Ciuti$^1$}
\affiliation{$^1$ Laboratoire Mat\'eriaux et Ph\'enom\`enes Quantiques, Universit\'e Paris Diderot, CNRS UMR 7162, Sorbonne Paris Cit\'e, 10 rue Alice Domon et Leonie Duquet 75013 Paris, France}
\affiliation{$^2$ ICTP, Strada Costiera 11, 34151 Trieste, Italy}
\affiliation{$^3$ NEST, Scuola Normale Superiore and Istituto Nanoscienze-CNR, I-56126, PISA, Italy}


\date{\today}

\begin{abstract}

We explore critical properties of  two-dimensional lattices of spins interacting via an anisotropic Heisenberg Hamiltonian and subject  to incoherent spin flips.
We determine the steady-state solution of the master equation for the density matrix via the corner-space renormalization method. We investigate the finite-size scaling and critical exponent of the magnetic linear susceptibility associated to a dissipative ferromagnetic transition. We show that the Von Neumann entropy increases across the critical point, revealing a strongly mixed character of the ferromagnetic phase. 
 Entanglement is witnessed by the quantum Fisher information which exhibits a critical behavior at the transition point, showing that quantum correlations play a crucial role in the transition.
 \end{abstract}

\pacs{}

\maketitle



\section{Introduction}

Quantum phase transitions~\cite{sachdev2001quantum} are fascinating critical phenomena affecting the nature of the quantum ground state in the thermodynamical limit, as a result of the competition between distinct physical contributions to the system Hamiltonian. If a quantum system is coupled to an external reservoir, dissipative phase transitions may take place. 
They appear due to the competition between the coherent Hamiltonian dynamics and dissipation processes \cite{PhysRevA.86.012116,PhysRevLett.110.257204, PhysRevX.6.031011,PhysRevX.5.031028,Exact_Bartolo}. In contrast to equilibrium critical phenomena at zero temperature, the physical properties do not depend on the Hamiltonian ground state, but instead on the steady-state density matrix of a master equation accounting for the dissipation. Nowadays, a large amount of experimental platforms are accessible to study the many-body properties of quantum driven-dissipative systems: arrays of optical microcavities \cite{RevModPhys.85.299}, superconducting circuits \cite{Tomadin:10,Houck2012,Houck_1D}, trapped ions \cite{Muller20121}, cold atoms in optical lattices \cite{RevModPhys.85.553}. 

Many fundamental questions associated to the physics of dissipative phase transitions in extended systems are open. In particular, the role and eventual emergence of criticality of quantum correlations as a function of the system spatial size is yet to be explored, in conjunction with the mixed character of the steady state. Moreover, the calculation of critical exponents for non-equilibrium phase transitions  is an outstanding problem. A first important step in this direction is provided by the Keldysh functional integral formalism. Using a renormalization group approach,  the emergence of an effective thermal phase transition has been predicted in several driven-dissipative models, such as lossy polariton condensates \cite{PhysRevLett.110.195301,PhysRevB.89.134310,PhysRevX.5.011017} and spin systems \cite{PhysRevB.93.014307}. However, up to date the Keldysh formalism has not allowed to study the role of quantum correlations and entanglement in dissipative phase transitions.

A physical system that has recently attracted interest for the presence of a genuine dissipative phase transition is a lattice of spins described by a Heisenberg XYZ Hamiltonian in presence of a dissipating environment, which tends to relax each spin into the $\vert s_z = -1/2\rangle $ state. 
In this model, the single-site Gutzwiller mean-field theory for the density matrix predicts a phase transition from a paramagnetic phase, where all the spins point along the $z$-axis, to a ferromagnetic phase, which presents a finite polarization in the $xy$ plane \cite{PhysRevLett.110.257204}. A recent study \cite{PhysRevX.6.031011} has shown that in 1D systems such transition disappears, while it should survive in two-dimensional lattices. The two-dimensional case is particularly challenging to describe and predict theoretically, because well-known techniques such as the Density Matrix Renormalization Group (DMRG)  \cite{RevModPhys.77.259} and Matrix Product Operators are more powerful for one-dimensional systems \cite{PhysRevLett.114.220601,PhysRevA.92.022116}.

In this article, we present a theoretical study of the  dissipative phase transition of the XYZ Heisenberg system in two-dimensional lattices (with periodic boundary conditions) by applying the recently developed corner-space renormalization method \cite{Physrevlett.115.080604}.
We determine the finite-size scaling of the magnetic susceptibility and show that its peak increases as a power law of the system size in agreement with the presence of a dissipative phase transition in 2D. A study of the Von Neumann entropy of the steady-state density matrix reveals that the ferromagnetic phase is a mixed state. 
Importantly, our results for the finite-size scaling of the quantum Fisher information \cite{PhysRevLett.72.3439,PhysRevA.85.022321,Hauke2016} show that a critical behavior of entanglement and quantum correlations occurs at the transition. The picture that emerges is quite intriguing and novel as compared to ordinary classical or quantum phase transitions. 
In the classical case, quantum fluctuations are irrelevant, while in a quantum phase transition the entanglement properties are critical \cite{RevModPhys.80.517}.
In this work, we show that a dissipative phase transition can share properties of {\it both} classical and quantum phase transitions. 

The paper is organized as follows. In Sec. \ref{sec:method}, we describe the dissipative spin model considered and we discuss the numerical methods used to calculate its steady-state properties. In Sec. \ref{sec:results}, we present the results obtained and finally, in Sec. \ref{sec:conclusions}, we draw our conclusions.

\section{Theoretical model and methods}\label{sec:method}

The model considered in this work is a 2D spin-$1/2$ lattice governed by the Heisenberg XYZ Hamiltonian ($\hbar = 1$)
\begin{equation}\label{eq:Hamiltonian}
\hat{H} = \sum_{\langle i,j \rangle} (J_x \hat{\sigma}_i^x \hat{\sigma}_j^x + J_y \hat{\sigma}_i^y \hat{\sigma}_j^y + J_z \hat{\sigma}_i^z \hat{\sigma}_j^z) \ ,
\end{equation}
where $\hat{\sigma}_i^{\alpha}$ ($\alpha = x,y,z$) are the Pauli matrices on the $i$-th spin of the system and the spin-spin coupling is between nearest neighbor sites. We will assume that the system is subject to a dissipative channel that leads to the following Lindblad
master equation for the steady-state density matrix

\begin{equation}\label{eq:MasterEquation}
\frac{d\hat{\rho}}{dt} = -i \left[ \hat{H}, \hat{\rho} \right] + \sum_j \mathcal{L}_j [\hat{\rho}] = 0,
\end{equation}
where the incoherent spin relaxation is described by 
\begin{equation}
\sum_j \mathcal{L}_j [\hat{\rho}]  = 
\gamma \sum_j \left ( \hat{\sigma}^{-}_j \hat{\rho} \hat{\sigma}^{+}_j  - \frac{1}{2} ( \hat{\sigma}^{+}_j \hat{\sigma}^{-}_j \hat{\rho}  +   \hat{\rho}  \hat{\sigma}^{+}_j \hat{\sigma}^{-}_j )  \right ),
\end{equation}
with $\hat{\sigma}_j^{\pm} = (\hat{\sigma}_j^{x} \pm i \hat{\sigma}_j^{y})/2$. If the spin-spin coupling between nearest neighbors is anisotropic in the $xy$ plane (i.e. if $J_x \ne J_y$ in Eq. \ref{eq:Hamiltonian}), such relaxation is at odds with the Hamiltonian. The competition between the coherent and incoherent dynamics can induce the dissipative phase transition from a paramagnetic state with no magnetization in the $xy$ plane to a ferromagnetic state with finite magnetization in the $xy$ plane. The determination of the steady-state density matrix for an open lattice system is a challenging task, whose complexity grows faster than what required to calculate the ground state of a close Hamiltonian system. In fact, the density-matrix lives in a space whose dimension is the {\it squared} of the Hilbert space.

In this work, we consider squared spin lattices consisting of $L \times L$ sites with periodic boundary conditions. The results for the lattices with $L= 2$ and $L=3$ (having Hilbert space of dimension $2^4 = 16$ and $2^9 = 512$ respectively) have been obtained via a brute-force temporal integration of the master equation in Eq. (\ref{eq:MasterEquation}). The results for the lattices with $L= 4$, $5$ and $6$ have been obtained using the corner-space renormalization method \cite{Physrevlett.115.080604}, an algorithm which allows us to target the relevant sub-space (corner of the Hilbert space) for the steady-state density matrix.  The convergence of the results is checked by increasing the dimension $\mathcal{M}_C$ of the corner space until a desired accuracy is reached. For instance, the convergence of the numerical results for the magnetization of the $6 \times 6$ lattice has been obtained with a corner space dimension of $\mathcal{M}_C \simeq 5000$, which is considerably smaller than the full Hilbert space, having dimension $2^{36} \simeq 6.8 \times 10^{10}$. The technical details of the corner-space renormalization method can be found in Ref. \onlinecite{Physrevlett.115.080604}. The temporal solution of the master equation in the corner space for the lattices with $L = 4, 5, 6$ has been obtained via the Montecarlo wavefunction method (averaging over quantum trajectories) \cite{PhysRevLett.68.580,PhysRevLett.70.2273,Molmer93,RevModPhys.70.101}. The number of quantum trajectories (up to $500$) has been chosen to achieve a relative error on the expectation values of the order of $1$  percent.

\section{Results and discussion}\label{sec:results}

To study the critical properties of such a class of systems, we will focus on the linear response of the system in presence of an applied polarizing field in the $xy$ plane, which modifies the Hamiltonian as follows:
\begin{equation}\label{eq:HamiltExtField}
\hat{H}_{ext}(h,\theta) = \hat{H} + \sum_j h \left( \cos(\theta) \hat{\sigma}_j^x + \sin(\theta) \hat{\sigma}_j^y \right).
\end{equation}
 Since the Hamiltonian is anisotropic, the induced in-plane magnetization per site $\vec{M}$ depends on the angle $\theta$ of the applied field. The linear response is summarized by the  susceptibility tensor
\begin{equation}\label{eq:SusceptibilityTensor}
\pmb{\chi} 
= \left( 
\begin{array}{cc}
\chi_{xx} & \chi_{xy} \\
\chi_{yx} & \chi_{yy} \\
\end{array}
\right)
\quad \quad \quad \textrm{with} \quad \chi_{\alpha \beta} = \frac{\partial M_\alpha}{\partial h_\beta} \Big{ |}_{h = 0},
\end{equation}
where  $h_x = h \cos(\theta)$ and $h_y = h \sin(\theta)$  and  the induced magnetization per site is 
\begin{equation} 
M_{\alpha} = \frac{1}{N} \sum_{j=1}^N \textrm{Tr}(\hat{\rho} \hat{\sigma}^{\alpha}_j) \, ,
\end{equation}
with $ \alpha = x,y$ and $N$ the number of lattice sites.

For the study of critical behavior, it is convenient to look at a single quantity, namely the 
angularly-averaged susceptibility
\begin{equation}\label{eq:averageSusceptibility}
\chi_{av}= \frac{1}{2 \pi} \int_0^{2 \pi} d\theta \, \frac{\partial{| \vec{M}(h,\theta)}|}{\partial h} \Big{ |}_{h = 0}\ ,
\end{equation}
where 
\begin{equation}
\frac{\partial{ |\vec{M}(h,\theta)}|}{\partial h} \Big{ |}_{h = 0} =   \left  | \left (
\begin{array}{c}
\chi_{xx} \cos(\theta) + \chi_{xy} \sin(\theta) \\
\chi_{yx} \cos(\theta) + \chi_{yy} \sin(\theta)
\end{array}
\right) \right | \,.
\end{equation}

\begin{figure}
\includegraphics[width=0.45\textwidth]{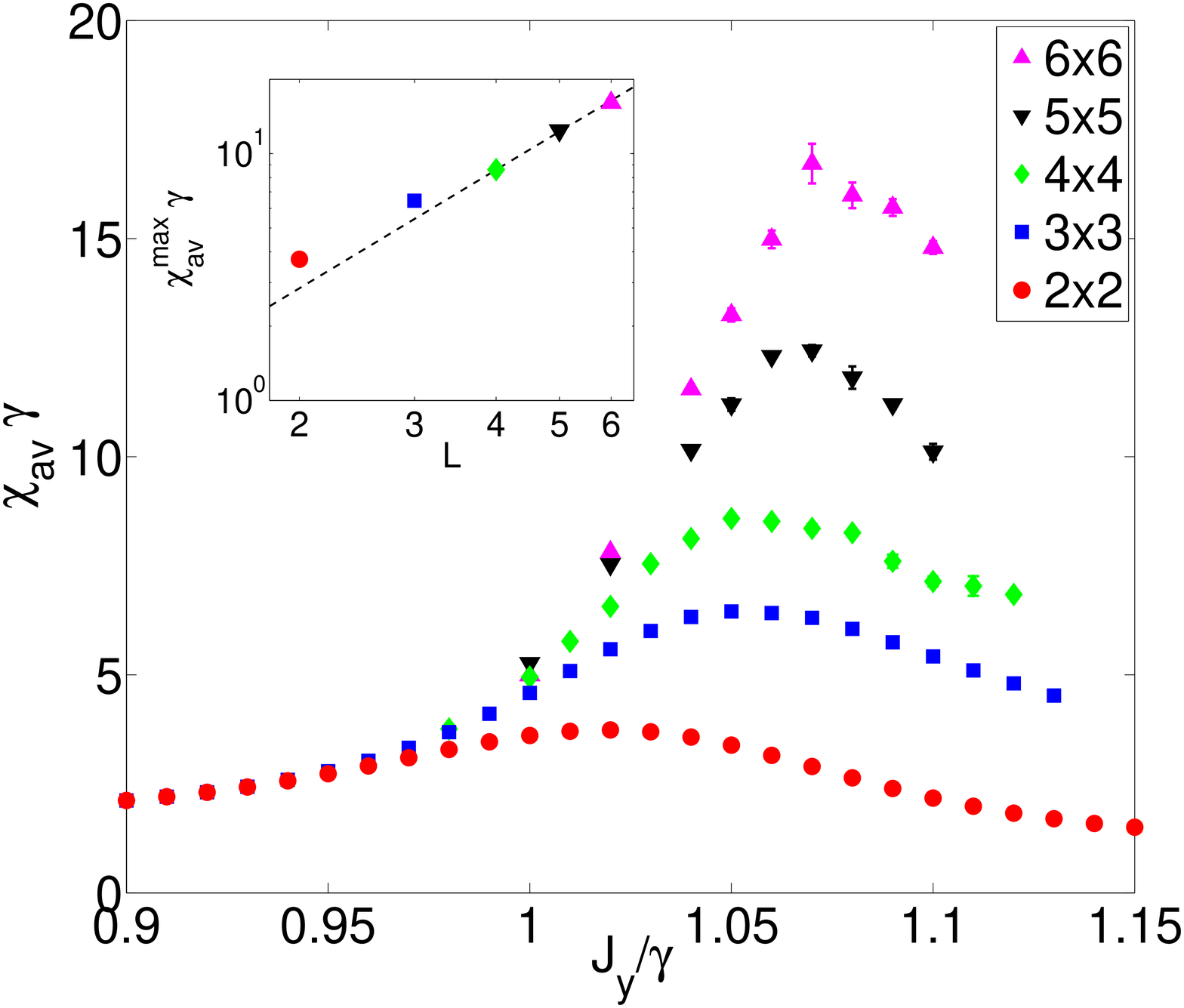}
\caption{Angularly-averaged magnetic susceptibility $\chi_{av}$ versus normalized coupling parameter $J_{y}/\gamma$ ($\gamma$ is the dissipation rate) for different sizes of the $L \times L$ lattice. The other coupling parameters are $J_x/\gamma = 0.9$ and $J_z/\gamma = 1$. Inset: maximum value $\chi_{av}^{max}$ of the susceptibility as a function of the size $L$ of the lattice. The dashed line is a power-law fit of the finite-size scaling. Error bars, when not shown, are smaller than the symbol size.}\label{fig:susceptibility}
\end{figure}

In Fig. \ref{fig:susceptibility}, we present the angularly-averaged susceptibility $\chi_{av}$ as a function of the normalized coupling $J_y/\gamma$  for squared lattices consisting of $L \times L$ sites with periodic boundary conditions (the other coupling parameters are  $J_x/\gamma = 0.9$ and $J_z/\gamma = 1$).  The convergence of the results with the dimension of the corner-space has been carefully controlled (the truncation errors fall within the statistical uncertainty arising from the Monte Carlo simulation). Each point  in Fig.  \ref{fig:susceptibility} has been determined considering four values of the applied field for each in-plane direction, in order to calculate accurately the linear susceptibility. For the $6 \times 6$ lattice in the critical region, the calculation of a single point in Fig. \ref{fig:susceptibility}  has required approximately one week of computation time on our 48-core computer cluster. 

The finite-size numerical results reported here are consistent with a critical behavior of the magnetic susceptibility \cite{domb1983phase} and allow us to obtain an estimate of the corresponding critical exponent. The magnetic susceptibility in Fig. \ref{fig:susceptibility} for the $L \times L$ lattices exhibits a peak of height $\chi_{av}^{max}(L)$ obtained for the coupling $J^{max}_{y}(L)$. As $L$ is increased, the peak becomes considerably sharper.    As shown in the inset, for the largest values of $L$, the peak height scales as a power law: in particular, the data for $L \ge 4$ are fitted by the  power law $\chi_{av}^{max}(L)  \propto L^{\kappa}$ with $\kappa = 1.59 \pm 0.10$, where the error correspond to one standard deviation.  To the best of our knowledge, this is the first numerical estimate of a critical exponent for a dissipative phase transition in two-dimensional spin lattices. It is interesting to notice that the numerical value $\kappa = 1.59 \pm 0.10$ is close to the corresponding critical exponent $7/4$ for the thermal transition of the 2D Ising model, but significantly deviates  from the value $2$, which is the mean-field critical exponent. However, in order to provide a more precise estimation of the critical exponent, in the future it will be necessary to perform simulations of larger lattices, which require further improvements of the present state-of-the-art methods. From the finite-size scaling of $J^{max}_{y}(L)$, we can also estimate a critical coupling  $J_{y}^{(c)}/\gamma = 1.07 \pm 0.02$. For a comparison, in the $4 \times 4$ cluster mean-field calculations in Ref. \cite{PhysRevX.6.031011}, the phase transition occurs at $J_y/\gamma \simeq 1.03$.

Cluster-mean field calculations predicted the presence of a second phase transition around $J_y/\gamma = 1.4$, although the location of this second critical point varies considerably for $2 \times 2$, $3\times 3$, $4 \times 4$ clusters \cite{PhysRevX.6.031011}. We performed exact calculations  (i.e., full Hilbert space) of the susceptibility for $2 \times 2$ and $3 \times 3$ lattices with periodic boundary conditions and for a $4 \times 4$ lattice (via the corner-space renormalization): for $J_y \gtrsim 1.4\gamma$, we do not observe any additional peak in the magnetic susceptibility (see Appendix). Our results with periodic boundary conditions suggest the absence of a second phase transition at large $J_y$, in disagreement with mean-field calculations with small clusters: further calculations for larger lattices, which would be thus less sensitive to specific boundary conditions, are therefore needed to fully understand the properties of the dissipative XYZ model in such a region of the phase diagram.

To study the mixed character of the steady-state density matrix, a useful quantity to evaluate is the 
Von Neumann entropy, defined as
\begin{equation}\label{eq:entropy}
S = - \textrm{Tr}(\hat{\rho}  \ln(\hat{\rho}) ) = - \sum_r p_r \ln(p_r)
\end{equation}
where $p_r$ are the eigenvalues of the density matrix $\hat{\rho} = \sum_r p_r | \Psi_r \rangle\langle \Psi_r |$. The calculation of the entropy is numerically much harder than that of the magnetization $\vec{M}$. Indeed, the convergence of $S$ versus the dimension $\mathcal{M}_C$ of the corner space is slower, since also the density-matrix eigenstates $| \Psi_r \rangle$ with small probability $p_r$  can give significant contributions to the sum in Eq. (\ref{eq:entropy}), due to the logarithmic term $\ln(p_r)$.
\begin{figure}
\includegraphics[width=0.45\textwidth]{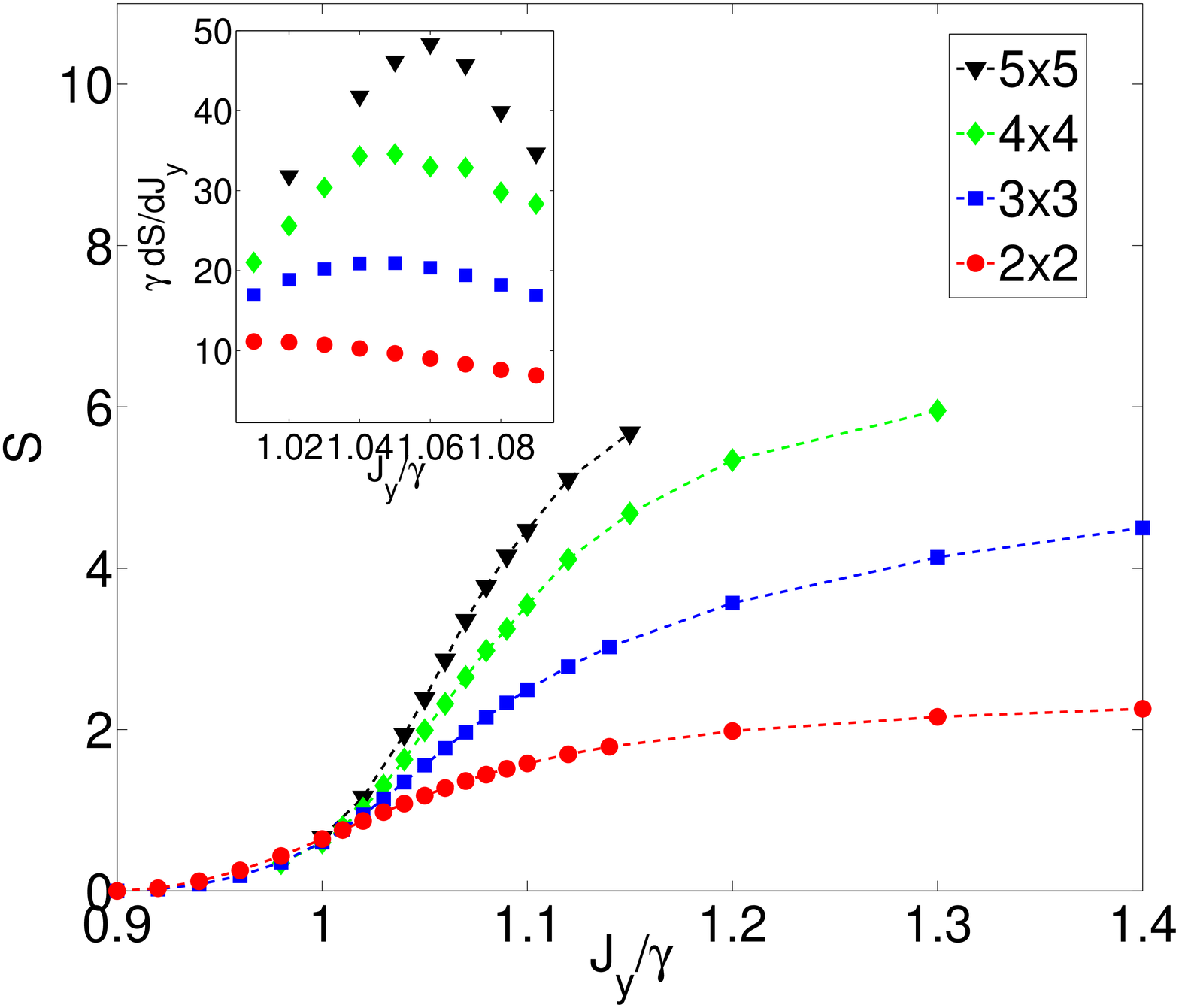} 
\caption{Von Neumann entropy $S$ as a function of the normalized coupling parameter $J_y/\gamma$ for different values of the size $L$ of the squared lattices. Same parameters as in Fig. \ref{fig:susceptibility}. Inset: the derivative of the entropy with respect to the coupling parameter $J_y$ . \label{fig:entropy} }
\end{figure}
Fig. \ref{fig:entropy} reports results for the entropy across the critical region. It is apparent that in the limit of isotropic system ($J_x = J_y = 0.9 \gamma$), the entropy tends to zero 
indicating a nearly pure state. Close to the critical point, the entropy sharply rises with a slope that increases with $L$. Finally, for large  $J_y/\gamma$, it saturates to a finite value which depends on the size of the lattice, indicating that the ferromagnetic phase is a strongly mixed state. In the inset of Fig. \ref{fig:entropy}, we present the numerical derivative of the entropy, showing a clear peak at the critical point, which becomes more and more pronounced by enlarging $L$. By fitting the maximum entropy derivative with a power-law (i.e., $\text{max} (\partial S/\partial J_y) \propto L^{\lambda}$) for $L \geq 3$, we get the estimate for the critical exponent $\lambda = 1.6 \pm 0.2$.The behavior of the Von Neumann entropy $S$ as a function of the coupling parameter $J_y$ in the XYZ model resembles the behavior of the entropy versus temperature in second-order thermal phase transitions.  However, in contrast to a classical transition, in the dissipative phase transition the ferromagnetic phase has larger entropy than the paramagnetic one.

An important question to address is whether quantum entanglement is present in the critical region. To this aim, we calculate the Quantum Fisher Information (QFI) of the mixed steady-state $\hat{\rho} = \sum_r p_r | \Psi_r \rangle\langle \Psi_r |$, which is defined as
\begin{equation}\label{eq:FisherDef}
F_Q = 2 \sum_{r,r^\prime} \frac{(p_r-p_{r^\prime})^2}{p_r+p_{r^\prime}} |\langle \Psi_r  |\hat{\mathcal{O}}| \Psi_{r\prime} \rangle|^2 \ ,
\end{equation}
where the sum includes only the terms with $p_r+p_{r\prime} > 0$. The operator $\hat{\mathcal{O}} = \sum_{j=1}^{N} \hat{\mathcal{O}}_j$ (where $N = L^2$ is the number of sites) is the sum of local hermitian operators $\hat{\mathcal{O}}_j$, whose spectrum width is 1 (the spectrum width is defined as the difference between the largest and the minimal eigenvalue). The value of the QFI is obtained by maximizing the expression in Eq. (\ref{eq:FisherDef}) over the possible operators $\hat{\mathcal{O}}$ with the properties above mentioned. The QFI has been used to study entanglement in quantum phase transitions in several systems at thermal equilibrium \cite{PhysRevA.80.012318, 1367-2630-16-6-063039, 0253-6102-63-3-279,Hauke2016}. Indeed, the QFI can be used as witness for multipartite entanglement:  if $F_Q/N > m$ then a quantum state possesses $m+1$-partite entanglement  \cite{PhysRevA.85.022321}. In particular, the inequality $F_Q / N > 1$ is a sufficient  condition to have bipartite entanglement \cite{PhysRevLett.102.100401}.

Since the order parameter of the transition in the dissipative XYZ model is the magnetization in the $xy$ plane, it makes sense to consider a set of operators $\hat{\mathcal{O}}$ similarly to Ref. \onlinecite{Hauke2016}. Namely, we evaluate $F_Q$ by considering $\hat{\mathcal{O}} =  \frac{1}{2} \sum_{j=1}^{N} \left( \cos(\tau) \hat{\sigma}_j^x + \sin(\tau) \hat{\sigma}_j^y \right)$ in Eq. (\ref{eq:FisherDef}) and by maximizing $F_Q$ with respect to the angle $\tau$. This procedure simplifies the numerical calculation and only produces an underestimation of the QFI. Consequently, if the inequality $F_Q / N > 1$ is satisfied, the bipartite entanglement is demonstrated a fortiori. The function $F_Q$ is directly accessible in our numerical calculations with the corner-space renormalization method. Indeed, one needs only to diagonalize the steady-state density matrix $\hat{\rho}$ in the corner-space to estimate $F_Q$.  However, similarly to calculation of the entropy, the convergence of the QFI  with respect to the corner-space dimension is more demanding than what needed for the susceptibility.  The criticality of the quantum correlations in the dissipative phase transition studied here can be characterized also with the calculation of other entanglement witnesses, such as the negativity \cite{PhysRevA.65.032314}. However, while the QFI can be directly calculated with the corner-space renormalization, the calculation of the negativity with our method is inefficient (see Appendix).

\begin{figure}
\includegraphics[width=0.45\textwidth]{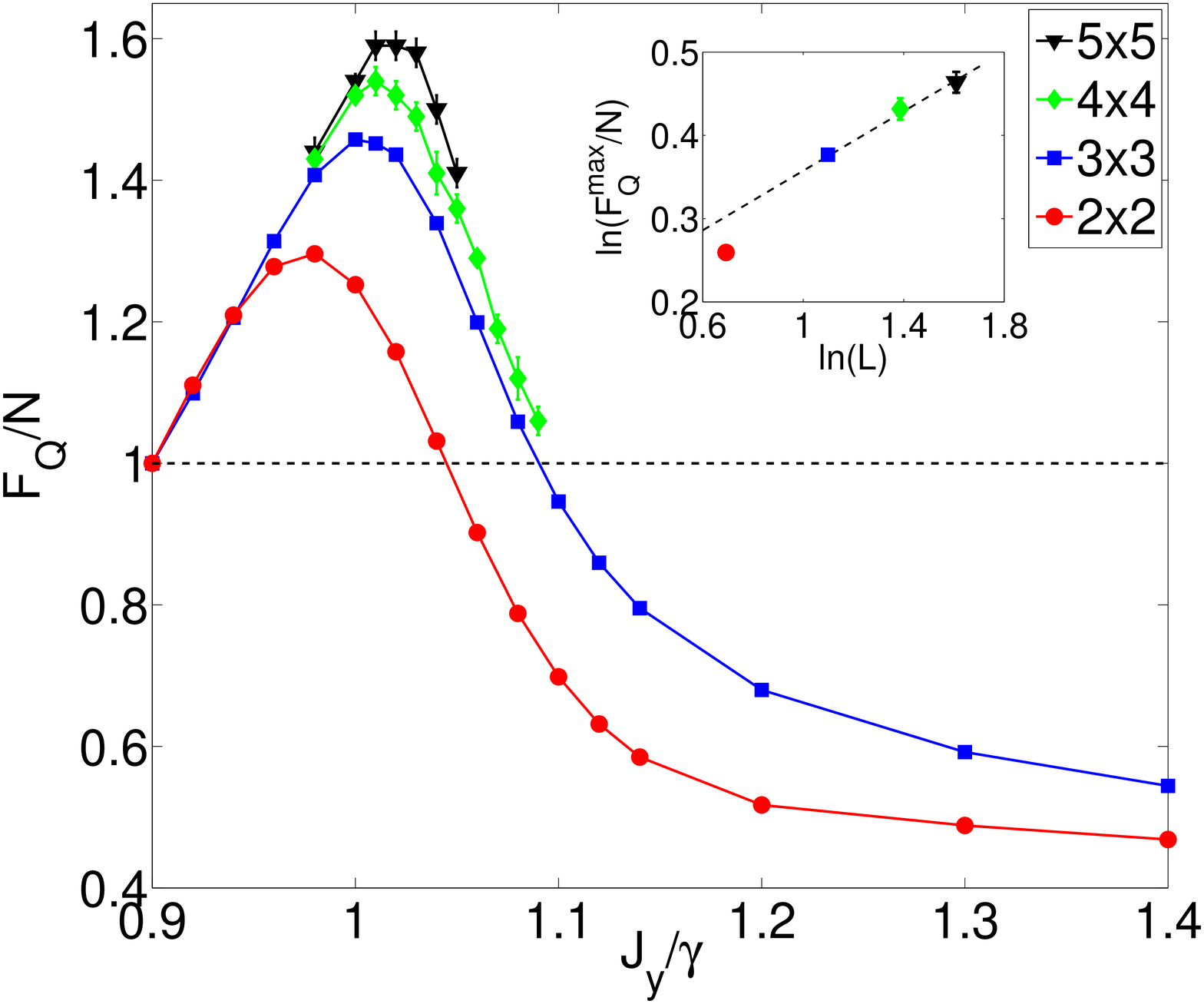}
\caption{Quantum Fisher Information $F_Q/N$ (normalized by the number $N = L^2$ of sites in the squared lattice) as a function of the normalized coupling parameter $J_y/\gamma$ for different sizes $L$.  Same parameters as in Fig. \ref{fig:susceptibility}. The inequality $F_{Q}/N > 1$ witnesses bipartite entanglement. Inset: maximum value of  $F_{Q}/N$ versus the lattice size $L$ (log-log scale) with a power-law fit (dashed line). }
\label{fig:FisherInfo}
\end{figure}
 
In Fig. \ref{fig:FisherInfo}, we show our results for $F_Q/N$ as a function of the coupling parameter $J_y$. It presents a maximum close to  $J_y^{(c)}$. In this regime, $F_{Q}/N > 1$ is sufficient to witness the presence of bipartite entanglement in the steady-state. Moreover, looking at the behavior of the maximum value of $F_Q/N$ versus $L$ of the lattice, we notice it increases for increasing $L$, although slower than the magnetic susceptibility peak. A power-law fit for $L \ge 3$ shows that the peak value $F^{max}_Q/N \propto L^{\eta}$ with
$\eta = 0.18 \pm 0.03$.

\section{Conclusions}\label{sec:conclusions}

We have explored theoretically a genuine dissipative phase transition of a two-dimensional spin lattice system described by an anisotropic XYZ Heisenberg Hamiltonian. By applying the corner-space renormalization method \cite{Physrevlett.115.080604} for the steady-state solution of the master equation, we have demonstrated that a critical behavior indeed emerges in two-dimensional lattices. By a finite-size scaling analysis of the magnetic susceptibility, we provided the first evaluation of the corresponding critical exponent. Our present work on the XYZ Heisenberg model shows that dissipative phase transitions share properties of both quantum and thermal phase transitions. Indeed, we have demonstrated that the Von Neumann entropy sharply increases across the transition, as it happens in thermal phase transitions.  Furthermore, the quantum nature emerges in the crucial role played by entanglement, as witnessed by the criticality of the quantum Fisher information. An interesting development is the study of different entanglement witnesses, to see how their critical properties may change. Future exploration of other physical models with different symmetries and dissipators is an intriguing perspective.

\acknowledgements
We acknowledge support from ERC (via  the Consolidator  Grant ``CORPHO'' No.  616233) and from   EU through contracts SIQS and QUIC.
We thank I. Carusotto, W. Casteels, T. Lacroix, D. Rossini and V. Savona for discussions and/or a critical reading of the manuscript.

\appendix
\section*{Appendix}\label{sec:appendix}

In this Appendix, we provide additional numerical results concerning the dissipative anisotropic XYZ Heisenberg model for two-dimensional lattices.

\begin{figure}[h!]
	\centering
	\includegraphics[width=0.42\textwidth]{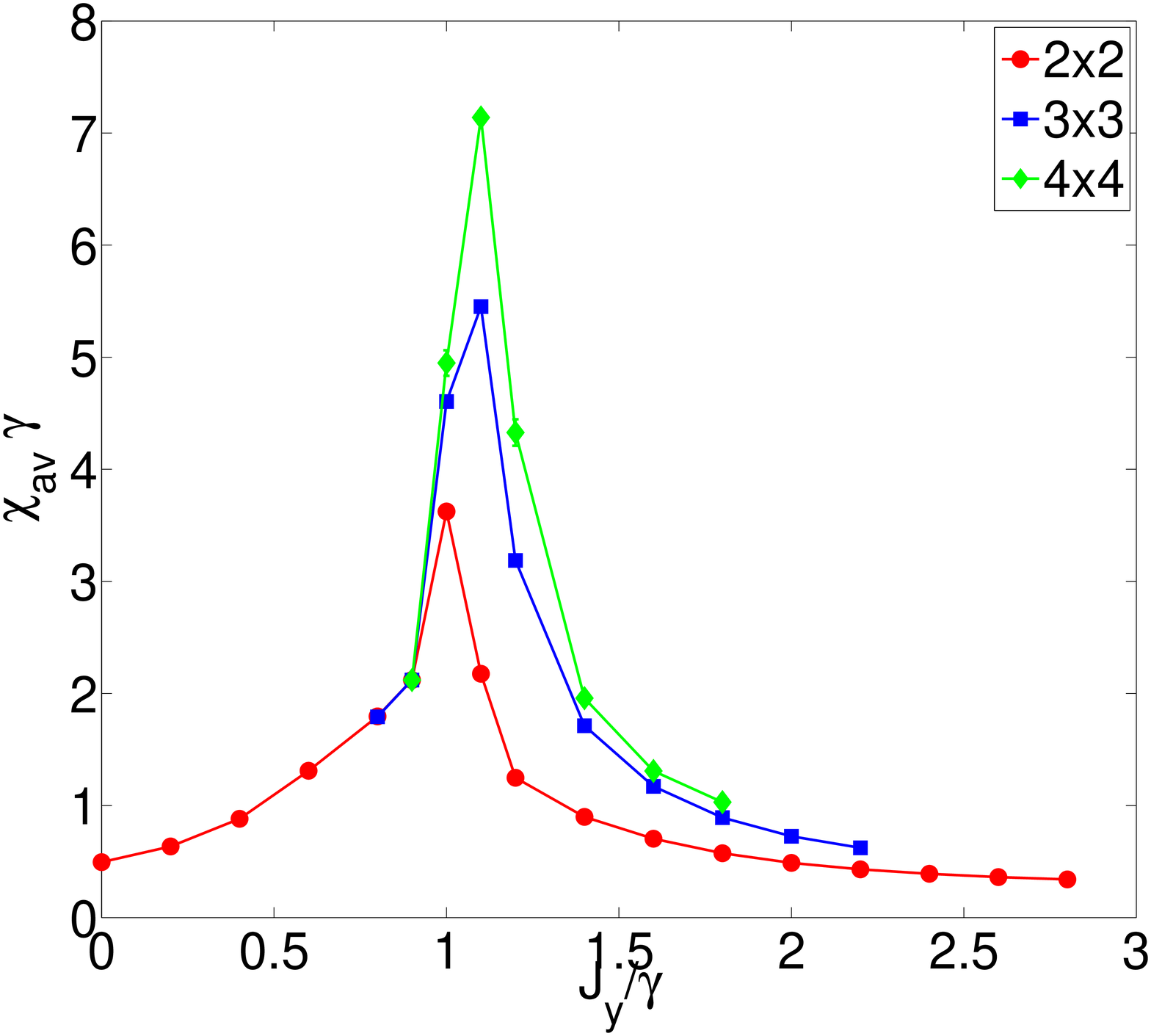}
	\caption{Angularly-averaged magnetic susceptibility $\chi_{av}$ versus normalized coupling parameter $J_{y}/\gamma$ for different sizes of the $L \times L$ lattice (periodic boundary conditions). Same parameters $J_x/\gamma$ and $J_z/\gamma$ as in Fig. \ref{fig:susceptibility}.}
	\label{Jy1.4}
\end{figure}

In Fig. \ref{Jy1.4}, we provide results for the angularly-averaged magnetic susceptibility on a wider range of values for $J_y/\gamma$. Notice that the results for $2 \times 2$ and $3 \times 3$ lattices  have been obtained considering the whole Hilbert space. The results for the $4 \times 4$ lattice have been calculated with the corner-space renormalization method. The susceptibility peak around $J_y \simeq \gamma$  is associated to the critical point comprehensively studied in Sec. \ref{sec:results}. We would like to point out that in the region of larger $J_y$, there is no additional susceptibility peak.\\

\begin{figure}
	\centering
	\includegraphics[width=0.42\textwidth]{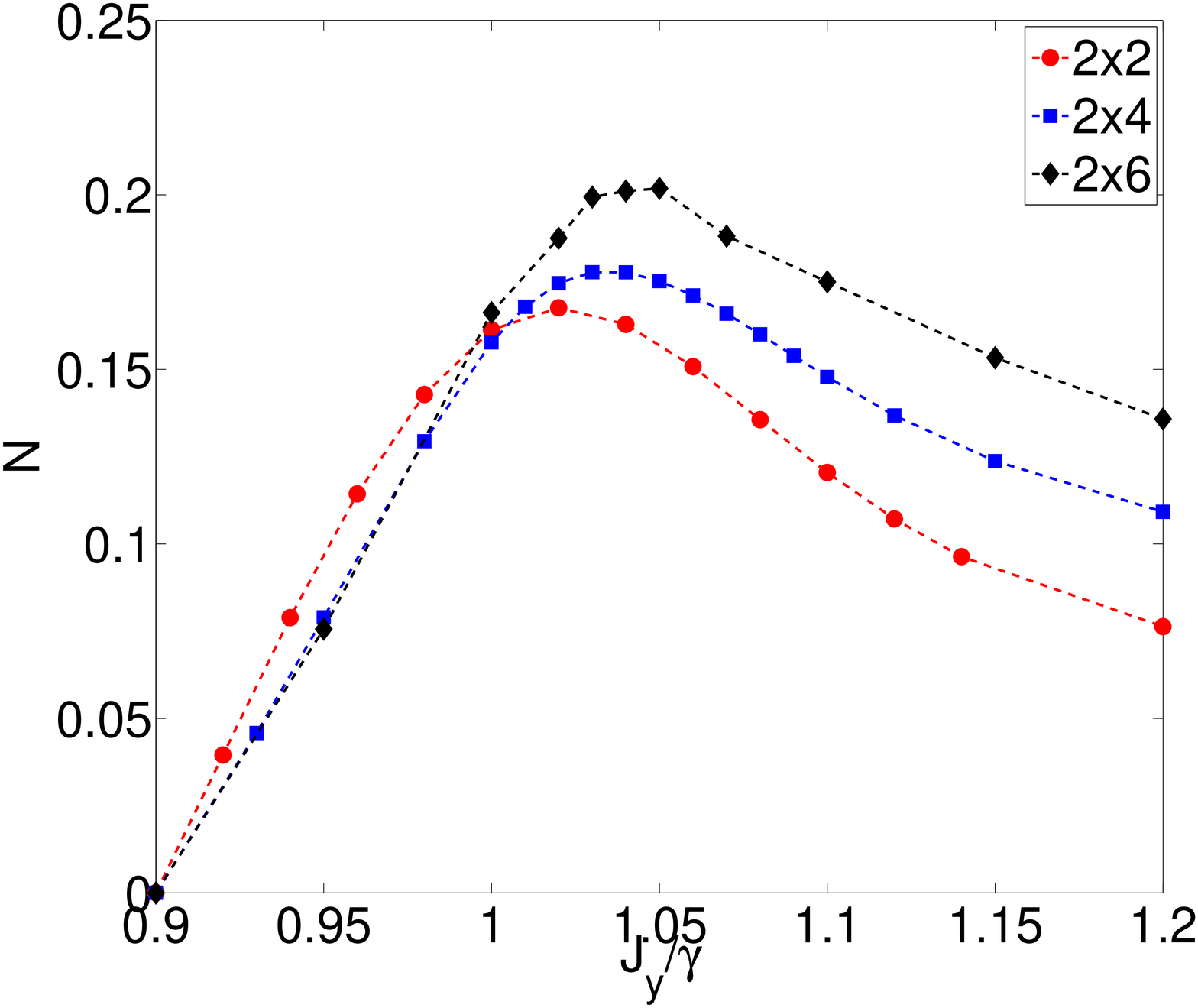}
	\caption{Entanglement negativity $\mathcal{N}$ vs. the normalized coupling parameter $J_{y}/\gamma$ for small lattices. Same parameters $J_x/\gamma$ and $J_z/\gamma$ as in Fig. \ref{fig:FisherInfo}.}
	
	\label{Negativity}
\end{figure}

In Fig. \ref{Negativity}, we show exact results (considering the whole Hilbert space) for the entanglement negativity\cite{PhysRevA.65.032314} $\mathcal{N}$   as a function of the normalized coupling parameter $J_{y}/\gamma$, for small rectangular lattices. By exact integration of the master equation in the full Hilbert space, we have found the steady-state density matrix of the $2 \times 2$, $4 \times 2$ and $6 \times 2$ lattices. We found that $\mathcal{N}(J_y)$ has a peak qualitatively similar to the Quantum Fisher Information. However, the negativity witnesses bipartite entanglement also in regions where the QFI does not, in particular above the critical point. Contrarily to the quantum Fisher information, the entanglement negativity cannot be calculated efficiently with the corner-space renormalization. Indeed, the calculation of $\mathcal{N}$ requires the diagonalization of the non-hermitian matrix obtained by partial transposition of the density-matrix with respect to one half of the lattice, an operation which enhances the truncation errors in the corner-space and makes unfeasible the calculation for larger lattices.

\bibliography{CriticalXYZ}

\end{document}